\documentclass[aps,prl,superscriptaddress,reprint,amsfonts,amsmath]{revtex4-1}
\usepackage[final]{graphicx}
\usepackage{comment}

\begin{document}
\title{Crystal Growth in Fluid Flow: Nonlinear Response Effects}
\date{\today}
\def\dlr{\affiliation{Institut f\"ur Materialphysik im Weltraum, Deutsches Zentrum f\"ur Luft- und Raumfahrt (DLR), 51170 K\"oln, Germany}}
\def\ubo{\affiliation{Experimentalphysik IV, Ruhr-Universit\"at Bochum, Universit\"atsstrasse 150, 44780 Bochum, Germany}}
\def\hhu{\affiliation{Department of Physics, Heinrich-Heine-Universit\"at D\"usseldorf, Universit\"atsstra\ss{}e 1, 40225 D\"usseldorf, Germany}}
\author{H.~L.~Peng}\dlr\ubo
\author{D.~M.~Herlach}\dlr\ubo
\author{Th.~Voigtmann}\dlr\hhu

\begin{abstract}
We investigate crystal-growth kinetics in the presence of strong shear
flow in the liquid, using molecular-dynamics simulations of a
binary-alloy model.
Close to the equilibrium melting point, shear flow always suppresses
the growth of the crystal--liquid interface. 
For lower temperatures, we find that the growth velocity of the crystal
depends non-monotonically on the shear rate.
Slow enough flow enhances the crystal growth, due to an increased
particle mobility in the liquid.
Stronger flow causes a growth regime that is nearly
temperature-independent, in striking contrast to what one expects from
the thermodynamic and equilibrium kinetic properties of the system, which
both depend strongly on temperature.
We rationalize these effects of flow on crystal growth as resulting from
the nonlinear response of the fluid to strong shearing forces.
\end{abstract}

\maketitle

Crystallization, a paradigmatic first-order phase transformation, is
of utmost importance in materials
science and engineering.
Many materials, for example most polymeric and metallic materials of daily life,
are produced from the liquid state as their parent phase, in the presence
of strong flow (e.g., in extrusion or casting processes).
Since crystal growth governs the evolution of the microstructure,
detailed knowledge of how crystallization is affected by the processing
conditions offers an effective
way to design and control material properties in applications
\cite{Herlach.2007,Kelton.2010,Yan.2016,Diao.2015,Anwar.2014}.
Flow effects have been studied extensively for metallic melts,
and they are in particular also relevant for soft materials, where typical flow
rates are of the order of typical structural relaxation times.
Yet, to understand the microscopic principles
of flow-induced changes to nucleation and crystal growth still presents
a challenge to statistical physics.

Crystallization from the melt consists of two stages: nucleation
of an initial crystalline seed, and subsequent growth.
Despite its simplifications, classical nucleation theory (CNT) continues
to be a useful reference for nucleation \cite{Bokeloh.2011}.
The effect of flow on nucleation as studied in simulation
\cite{Richard.2015,Wu.2009,Cerda.2008,Mokshin.2008,Shao.2015}
can reasonably well be understood
by accounting for the flow-induced changes in the near-equilibrium
thermodynamic quantities of CNT
\cite{Blaak.2004,Mokshin.2013,Mura.2016,Lander.2013}.

Crystal growth on the other hand needs to be discussed as a
non-equilibrium process that is strongly
influenced by external driving forces \cite{Dorosz.2016}.
Once an initial crystal has formed, in a sheared liquid
crystal growth and flow-induced erosion compete and lead to a
non-equilibrium coexistence that depends on temperature and flow rate
\cite{Butler.2002}.
It has already been emphasized that a near-equilibrium thermodynamic
description of flow-modified growth kinetics is not viable
\cite{Butler.2002,Butler.2003,Butler.2003b}.
It is therefore much less clear, how crystal growth changes under strong
fluid flow, and which are the governing microscopic processes.

Here we present molecular-dynamics (MD) simulations of crystal growth
in a homogeneously sheared fluid, over a wide range of temperatures and
shear rates.
We argue that the nonlinear response of the fluid
to the shearing force is a relevant microscopic dynamical process
that determines a set of qualitatively different growth regimes.
Strikingly, we find that the growth velocity of the crystal in a deeply
undercooled fluid is a non-monotonic function of the shear rate. After
an initial strongly temperature-dependent increase, a regime of intermediate
shear rates appears where the crystal grows with a velocity that is nearly
temperature independent. It is rationalized as the result of strong shear thinning
of the undercooled fluid.
This nonlinar-response effect has so far, to our knowledge, been neglected
in the modeling of solidification processes. It is achieved once shear-induced
``surface erosion'' and structural relaxation of the viscoelastic melt
compete. This regime of flow rates is of particular relevance for soft
materials, but can in principle also be reached for sufficiently undercooled
metallic melts.

\begin{figure}
\includegraphics[width=.9\linewidth]{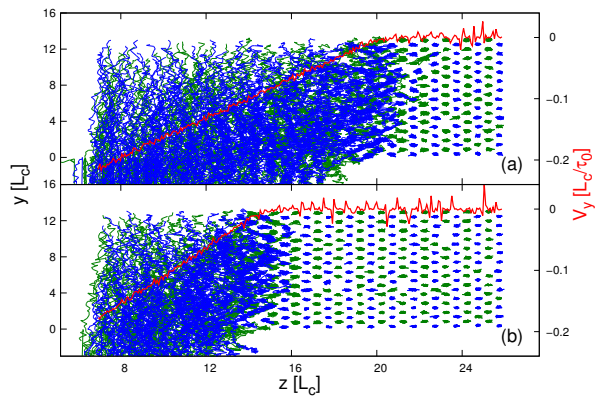}
\caption{ \label{fig:snapshot}
Snapshots of the simulation of crystal growth along the $z$ direction, in a
fluid sheared in the $y$--$z$ plane, for two different times. Units are
fixed by the microscopic relaxation time $\tau_0$ and the lattice constant
$L_c$ ($\tau_0=0.2\,\text{ps}$ and $L_c=2.897\,\text{\AA}$ for Al-Ni \protect\cite{Mishin.2002}).
Blue and green lines indicate the positions of Ni and Al atoms, respectively,
over a small time window. Red lines indicate the average velocity of the
atoms along the $z$ direction, highlighting the development of a homogeneous
linear shear profile in the fluid region.
}
\end{figure}

We performed molecular-dynamics (MD) simulations of a binary mixture
that crystallizes into a $B_2$ structure (CsCl lattice).
To provide a specific reference point, we use an embedded-atom
model of the intermetallic compound
$\text{Al}_\text{50}\text{Ni}_\text{50}$ \cite{Mishin.2002},
a material that is used in many applications, e.g., for turbines in
aeroplanes or power stations. $\text{Al}_\text{50}\text{Ni}_\text{50}$
melts congruently, so that crystal growth can be studied without any
constitutional effects.
The model has also been extensively studied in previous MD simulations,
investigating
its liquid dynamics \cite{Das.2005}, glass-forming ability \cite{Tang.2013}, quiescent
crystal growth \cite{Kerrache.2008,Kuhn.2013,Tang.2013}, and disorder trapping \cite{Zheng.2012}.
Using $N=27040$ particles in a box of average dimensions
$L_x:L_y:L_z=1:1:6.4$ and $L_x\approx38\,\text{\AA}$ (employing periodic
boundary conditions in all Cartesian directions), seeded with an initial
crystal that is surrounded by liquid regions,
we study crystal growth along the normal of the (100) face
(along the $z$-direction).
The system is first prepared in its $B_2$ state and relaxed to the target
temperature. Next, particles in the central third of the box ($L_z/3
\approx81\,\text{\AA}$) are fixed, and the surrounding system is molten at
$T=3000\,\text{K}$ before it is brought to the desired temperature again.

Simple shear flow parallel to the interface is imposed by assigning a fixed
center-of-mass velocity to small fluid layers (width $12\,\text{\AA}$)
at the $z$-boundaries of the
simulation box, while keeping a layer of $10\,\text{\AA}$ fixed in the
crystalline center. After an initial transient (of about $100\,\text{ps}$),
a homogeneous linear velocity profile develops in the liquid, as shown in
Fig.~\ref{fig:snapshot}. The shear rate, $\dot\gamma$,
is extracted from the averaged velocity, $\dot\gamma=\partial_zv_y$.
Data are analyzed in small time
windows during which the shear rate does not change appreciably.
To release latent heat as the crystal grows,
we employ a profile-unbiased thermostat \cite{Evans}
in slabs (of width $4\,\text{\AA}$) parallel to the interfae \cite{Monk.2010}.
A barostat along the growth direction keeps $p_z=0$ and
compensates for the volume expansion
during the crystallization process.

The time-dependent crystal--liquid interface position $z_I$ is obtained from
both the local fluid velocity,
as the point where $|v_y(z_I)|<0.05\,\text{\AA}/\text{ps}$,
and from the crystalline order-parameter field, $\Psi(z)$, as the point
where $\Psi(z_I)>0.01$. Here, $\Psi(z)$ measures four-fold symmetry in
the plane perpendicular to the growth direction:
\begin{equation}
  \Psi(z)=\langle\sum_i\delta(z_i-z)\frac1{M(M-1)}
  \sum_{j\neq k}\cos(8\theta_{ijk})\rangle\,,
\end{equation}
where the sums over $j$ and $k$ extend over the nearest-neighbor atoms
of particle $i$ (defined as those $M$ atoms whose relative distance is less
than that of the first minimum in the liquid-state pair distribution function),
and $\theta_{ijk}$ is the angle in the $x$-$y$-plane between
the distance vectors $\vec r_{ij}$ and $\vec r_{ik}$.
The interface positions obtained by both methods differ by about
$5\,\text{\AA}$, indicating a hydrodynamic slip length, but give consistent
growth velocities.

\begin{figure}
\includegraphics[width=.9\linewidth]{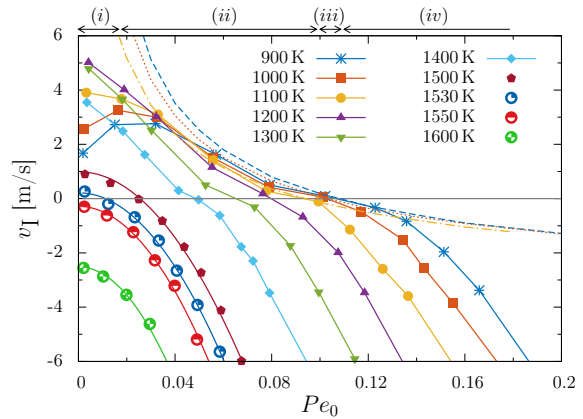}
\caption{ \label{VI-u}
Crystal--fluid interface velocity $v_I$ as a function of shear rate $Pe_0=\dot\gamma\tau_0$ for different temperatures $T$ as labeled. Labels (i) to (iv) indicate for $T=900\,\text{K}$ the different growth regimes discussed in the text.
Dashed lines are estimates from Eq.~\eqref{vi-hi} for $T\le1100\,\text{K}$.
}
\end{figure}

The temperature- and shear-rate dependent growth velocity $v_I(T;\dot\gamma)$
of the crystal is shown in Fig.~\ref{VI-u} as a function of shear rate
$\dot\gamma$. Here and in the following, we use the time scale of atomic
vibrations, $\tau_0\approx0.2\,\text{ps}$ \footnote{The time scales of
single-particle vibrations, $\tau_0$,
and that of collective structural relaxation, $\tau$,
have been determined from the intermediate scattering
functions in the quiescent fluid, at wave vectors corresponding to
typical interparticle distances.} to define a dimensionless shear rate,
the (bare) P\'eclet number $Pe_0=\dot\gamma\tau_0$.
We briefly remark on the behavior close to the equilibrium
melting temperature ($T_m\approx1540\,\text{K}$ in our model).
Here, shear flow always suppresses crystal growth, and
the growth velocities $v_I$ closely follow a quadratic dependence on
$\dot\gamma$.
The zeroes of the $v_I$-versus-$\dot\gamma$ curves identify
a non-equilibrium crystal--fluid coexistence (at a shear rate
$\dot\gamma_\text{coex}(T)$) that has been discussed earlier
\cite{Butler.2002,Butler.2003,Butler.2003b}.
Crystal growth velocities close to $T_m$ are usually expressed in terms
of thermodynamic free-energy barriers, and it is tempting to relate
the quadratic decrease of $v_I$ to a
shifted effective chemical-potential difference between the liquid and
the crystal (and hence to a process-dependent effective undercooling
\cite{Herlach.2014}).
However,
the shear-induced change in the strained crystal's free energy
is much too small to quantitatively explain the simulation data
\cite{Butler.2003}.

For deeper undercooling, in particular, the situation is far more complex:
the growth rate displays a non-monotonic dependence on the shear rate at
any fixed temperature, with a maximum at a non-zero intermediate rate,
and an inflection point around $\dot\gamma_\text{coex}$. Between the
maximum and the coexistence point, the temperature dependence of the
growth rate is remarkably weak, i.e., the interface velocity is described
well by a temperature-independent master curve $v_I(\dot\gamma)$ in
an intermediate shear-rate regime.
This is unexpected from the point
of view of thermodynamics (where Boltzmann factors imply a temperature
dependence when keeping other parameters fixed),
or from the strong slowing down of the
transport kinetics in the fluid that occurs upon cooling.


To understand the mechanisms responsible for the non-monotonic dependence
of the growth velocity on the shear rate, let us divide the $v_I$-versus-$\dot\gamma$ curves into
four regimes, as labeled in Fig.~\ref{VI-u}:
(i) for slow shear, the suppression of
growth due to shear that is observed close to $T_m$,
changes to an enhancement below some
temperature ($T\approx1000\,\text{K}$ in the figure). (ii) At intermediate
shear rates, $v_I$ decreases towards zero with increasing $\dot\gamma$.
The curves for
$T\lesssim1200\,\text{K}$ approach a $T$-independent master curve for
$0.03\lesssim Pe_0\lesssim0.1$
($0.15\,\text{ps}^{-1}\lesssim\dot\gamma\lesssim0.5\,\text{ps}^{-1}$).
The interplay between (i) and (ii) causes the maximum in the growth velocity
for a finite shear rate, visible at the lowest temperatures shown
in Fig.~\ref{VI-u}(a).
(iii) The coexistence regime around
$v_I(\dot\gamma_\text{coex})=0$ becomes broader in the
sense that the depedence of the growth velocity on shear rate becomes
weak, i.e., near-stationarity of the interface is achieved for a wider
range of shear rates.
(iv) At larger shear rates,
the crystal shrinks again.

\begin{figure}
\includegraphics[width=.9\linewidth]{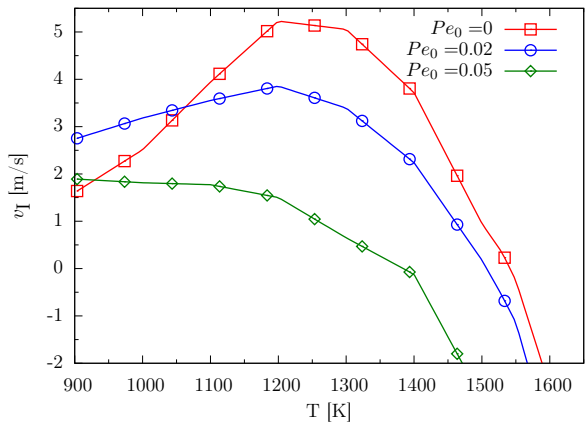}
\caption{\label{VI-T}
Interface velocity $v_I$ as a function of temperature $T$, for constant
shear rates given by bare P\'eclet numbers
$Pe_0=\dot\gamma\tau_0=0$ (squares), $0.02$ (circles), and $0.05$ (diamonds).
}
\end{figure}

For the explanation of regime (i) of Fig.~\ref{VI-u},
recall that
the undercooled liquid dynamics is characterized by slow structural relaxation:
density fluctuations decay on
a time scale $\tau\gg\tau_0$.
With decreasing temperature, $\tau$ increases much more rapidly
than expected from the high-temperature Arrhenius behavior.
Hence, the mobility of atoms in the liquid drops sharply, and
the crystal-growth mechanism in the quiescent liquid changes
from thermodynamically limited around $T_m$ to kinetically
limited at $T\ll T_m$.
As shown in Fig.~\ref{VI-T}, the resulting growth velocity
$v_I(T;\dot\gamma\!=\!0)$ exhibits
a maximum as a function of temperature (around $T=1200\,\text{K}$ in our
simulation) 
\cite{Tang.2013,Wang.2014}.

\begin{figure}
\includegraphics[width=.9\linewidth]{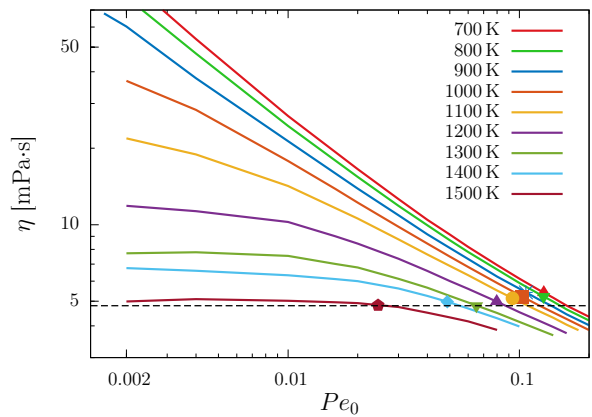}
\caption{ \label{VI-eta}
  Bulk fluid viscosity $\eta(\dot\gamma)$ as a function of shear rate,
  $Pe_0=\dot\gamma\tau_0$. Symbols mark the points where for each temperature,
  $\dot\gamma=\dot\gamma_\text{coex}$, i.e., the points where $v_I=0$.
  A horizontal dashed line indicates the equilibrium melting-point viscosity $\eta_m$.
}
\end{figure}

Structural relaxation speeds up in the sheared fluid when the
(dressed) P\'eclet number $Pe=\dot\gamma\tau$ is of the order
of the strain required to break typical nearest-neighbor cages,
$\gamma_c\approx10\%$. This is a
well-known nonlinear-response effect in metallic melts
\cite{Guan.2010} 
and most viscoelastic fluids \cite{Voigtmann.2014} that gives rise to
shear thinning -- a pronounced decrease in the fluid's shear viscosity with
increasing shear rate, as shown in Fig.~\ref{VI-eta} for our system.
All data shown in Fig.~\ref{VI-u} for $T\lesssim1100\,\text{K}$
correspond to $Pe\gtrsim\gamma_c$.
As a result, particle mobility in the fluid is enhanced by the flow, and the
growth velocity $v_I$ increases with increasing $\dot\gamma$ initially
for those $T$ where the equilibrium growth is limited kinetically.

Regime (ii) is essentially $T$-independent; this is clearly seen in
Fig.~\ref{VI-T}, where the curve for $Pe_0=0.05$ approaches
a constant for low temperatures. As opposed to regime (i),
$v_I(\dot\gamma)$ now decreases with increasing shear rate, which indicates
that a qualitatively different process limits the growth.
Hydrodynamic momentum
transport across the interface region is governed by the rate
$ r_+\sim{\eta}/{L^2\rho} $,
where $\rho$ is the fluid mass density and $L$ the interface
width ($L\approx10\,\text{\AA}$ in our simulations for all state points).
Note that at $T_m$, one gets $r_+\approx1/\tau_0$, the natural scale
for the rate of momentum transport in the crystal.
Assuming
that the rate $r_+$ limits the attachment of atoms to the interface
and hence the growth, and that it balances the detachment
rate at the non-equilibrium coexistence point, we get
\begin{equation}\label{vi-hi}
  v_I(\dot\gamma)\sim
  v_0\cdot(\eta(\dot\gamma)-\eta(\dot\gamma_\text{coex}))\,\tau_0/L^2\rho \end{equation}
where $v_0$ is a velocity scale set by the thermodynamic features of the system ($v_0=\mathcal O(1\,\text{m}/\text{s})$ in our simulation).
In Eq.~\eqref{vi-hi} we have accounted for the fact that the viscosity
$\eta(\dot\gamma)$ depends sensitively on the shear rate in the
shear-thinning regime (ii). At the same time,
structural relaxation in the fluid becomes nearly
$T$-independent, since it is dominated by shear and thus $1/\dot\gamma$ is
the only relevant time scale. The fluid then flows plastically,
i.e., at the expense of a nearly constant yield stress $\sigma_y$. As
a result, also the shear viscosity $\eta(\dot\gamma)$ entering
Eq.~\eqref{vi-hi} is nearly temperature-independent.
Expression \eqref{vi-hi}
is shown for the lowest three temperatures in Fig.~\ref{VI-u}
(dashed lines),
evaluated using the viscosity of the bulk fluid (see below).
It gives a reasonable
qualitative account for $v_I(\dot\gamma)$ in the range $0.04
\lesssim Pe_0\lesssim0.1$. For comparison, a thermodynamic
argument based on an effective free-energy barrier would
not account for the weak $T$-dependence we observe.
Equation~\eqref{vi-hi} naturally explains that in regime (ii), the growth
velocity
decreases with increasing shear rate: for true yield-stress flow,
$\eta(\dot\gamma)\sim\sigma_y/\dot\gamma$, and thus $v_I\sim1/\dot\gamma$
up to a constant and weakly dependent on temperature.

Equation~\eqref{vi-hi} implies that the fluid viscosity determines the non-equilibrium
coexistence between the sheared liquid
and the crystal, region (iii).
To corroborate this argument, we show in Fig.~\ref{VI-eta} the shear-rate
dependent viscosity of the bulk liquid, as determined from separate
MD simulations (in a system
of $N=5000$ particles following the SLLOD equations of motion \cite{Evans}).
Despite the fact that the bulk-liquid viscosity drops by more than
an order of magnitude over the range of shear rates we investigate,
the viscosity
at the coexistence point (marked by symbols in Fig.~\ref{VI-eta}) only varies
by about $20\%$ around the equilibrium melting-point viscosity $\eta_m$.
Thus,
\begin{equation} \eta(\dot\gamma_\text{coex})\approx\eta_m\,. \end{equation}
This implies that in regime (iii), the dependence
of $v_I(\dot\gamma)$ on the flow rate is weak, as indeed observed
in Fig.~\ref{VI-u} for $T\lesssim1100\,\text{K}$.
It indicates a rate-controlled non-equilibrium coexistence that is
attributed to the non-Newtonian fluid behavior, and that
is different from a thermodynamical balance.

\begin{figure}
\includegraphics[width=.9\linewidth]{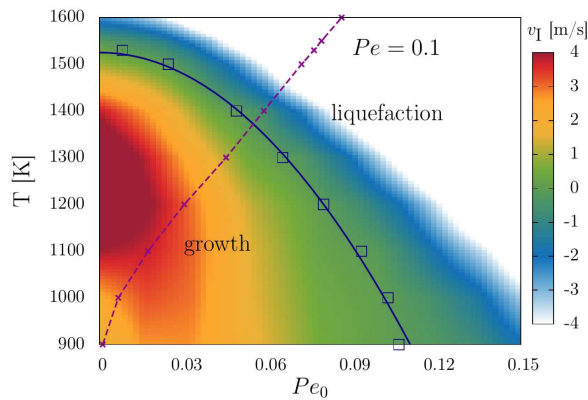}
\caption{ \label{T-gamma}
  Nonequilibrium crystallization diagram for the sheared $\text{Al}_\text{50}\text{Ni}_\text{50}$ fluid and $\text{B}_\text{2}$ crystal in the temperature--shear-rate plane. Colors indicate the growth velocity as labeled. The red line is a quadratic fit for the coexistence line.
  The dashed line indicates the cross-over from linear response to shear
  (left) to th regime of nonlinear-response flow (right).
}
\end{figure}

A coexistence point is intuitively expected
since shear-induced erosion of particles from the interface
eventually becomes strong enough to suppress any attachment.
In agreement with earlier simulations \cite{Butler.2002}
and a recent model \cite{Ramsay.2016},
a non-equilibrium
phase diagram is obtained in the $T$-$\dot\gamma$-plane,
shown in Fig.~\ref{T-gamma}. A quadratic fit
(red line in Fig.~\ref{T-gamma}), gives a good description
of $T_\text{coex}(\dot\gamma^2)$, extrapolating to $T_\text{coex}(0)
\approx1531\,\text{K}$ in reasonable agreement with the value of 
$T_m$ reported for the model before \cite{Kerrache.2008}.
The surface-erosion picture is consistent with a weak $\dot\gamma$-dependence
of $v_I$ close to $\dot\gamma_\text{coex}$:
As long as shear-induced changes in the thermodynamic forces are
negligible, any fast enough shear rate will be sufficient to erode
fluctuations in the interface.

Only in regime (iv), effective thermodynamic forces again
dominate the crystallization process; they always lead to shear-induced
melting of the crystal. To the strongly sheared liquid, an increased
effective temperature $T_\text{eff}(\dot\gamma)\ge T$ can be assigned
\cite{Berthier.2002},
leading to a reduced effective undercooling that controls the growth velocity.
Indeed, our data in regime (iv) closely follows a single
$\eta$-dependent curve, where the viscosity is a proxy
for the fluid's effective temperature.


In conclusion, the growth velocity of a crystal is a non-monotonic function
of the shear rate at fixed temperature. Flow initially enhances crystal
growth at sufficiently strong undercooling, due to the enhanced particle
mobility in the fluid. In the presence of stronger flow, the hydrodynamics
of surface erosion causes a decrease in the growth velocity as a function
of shear rate that is insensitive to temperature. This non-monotonic
dependence is rationalized as the result of
the pronounced nonlinear-response behavior of the non-Newtonian undercooled
fluid.

The strain supported by nearest-neighbor cages marks
the cross-over from the regime of small
shear rates to the nonlinear-response dominated regime.
This cross-over is reached once the dressed P\'eclet number
(formed with the structural-relaxation time rather than the timescale
of atomic vibrations) reaches $Pe\approx0.1$.
The line corresponding to $Pe=0.1$ is
shown in Fig.~\ref{T-gamma} (dashed line).

The mechanism we propose is very generic since it rests solely
on the fact that the liquid becomes shear
thinning. This is the case for most undercooled (viscoelastic) fluids,
including colloidal suspensions, soft materials, as well as metallic melts.
Our results point out that nonlinear-response effects in the undercooled
fluid likely should be taken into account in effective coarse-grained models
of rapid solidification \cite{Lee.2014,Shevchenko.2015,Rojas.2015}.
It is difficult to assess the actual shear rates in experiments on metallic
melts.
Still, an enhancement of crystal growth by
(convective) flow has been reported in experiments
combining high-speed imaging, electromagnetic levitation
and microgravity conditions \cite{Reutzel.2007,Binder.2014,Herlach.2014}
to determine crystal growth velocities of alloys in the presence of convective
flow on a mesoscopic scale.
For the change
in microscopic growth kinetics that occurs at $Pe\approx0.1$,
the relevant shear rate is a local
velocity gradient in front of the interface, extending over about $5$ to
$10$ atomic layers in our model.
Local fluid flow velocities
can be in excess of tens of $\text{m}/\text{s}$, indicating that the nonlinear
regime to right of the cross-over curve in Fig.~\ref{T-gamma} might be
reached in melts that can be undercooled sufficiently far.

The temperature-independent growth regime
in particular
is a signature of plastic yield-stress flow.
It suggests the use of controlled-flow conditions in applications where the speed of crystallization needs to be adjusted separately from thermal control.
We expect this mechanism to apply also to
more complex systems and geometries, for example in the presence of concentration gradients \cite{Yang.2011}.
In fact, $\text{Al}_\text{50}\text{Ni}_\text{50}$ is already an example where the quiescent growth mechanism is an intricate combination of attachment and intra-layer delayed reorganization \cite{Kuhn.2013}. Although we observe disorder trapping at the highest growth velocities, similar to the quiescent case at strong undercooling \cite{Hartmann.2009,Zheng.2012}, such details appear to leave the qualitative appearance of the different growth regimes as a function of shear rate unchanged.
Our findings will be essential to further explore unusual growth kinetics
observed in metallic materials, e.g., in Al-rich Al-Ni alloys where the
growth velocity decreases with increasing driving force and in the presence
of forced convection \cite{Lengsdorf2010}.

\begin{acknowledgments}
We acknowledge funding from Deutsche Forschungsgemeinschaft (DFG)
through grant He~1601/26.
We thank Peter Harrowell, J\"urgen Horbach and Jianrong Gao for discussions and valuable comments, and J\"urgen Brillo for a careful reading of the manuscript.
\end{acknowledgments}

\bibliography{lit}

\begin{thebibliography}{42}%
\makeatletter
\providecommand \@ifxundefined [1]{%
 \@ifx{#1\undefined}
}%
\providecommand \@ifnum [1]{%
 \ifnum #1\expandafter \@firstoftwo
 \else \expandafter \@secondoftwo
 \fi
}%
\providecommand \@ifx [1]{%
 \ifx #1\expandafter \@firstoftwo
 \else \expandafter \@secondoftwo
 \fi
}%
\providecommand \natexlab [1]{#1}%
\providecommand \enquote  [1]{``#1''}%
\providecommand \bibnamefont  [1]{#1}%
\providecommand \bibfnamefont [1]{#1}%
\providecommand \citenamefont [1]{#1}%
\providecommand \href@noop [0]{\@secondoftwo}%
\providecommand \href [0]{\begingroup \@sanitize@url \@href}%
\providecommand \@href[1]{\@@startlink{#1}\@@href}%
\providecommand \@@href[1]{\endgroup#1\@@endlink}%
\providecommand \@sanitize@url [0]{\catcode `\\12\catcode `\$12\catcode
  `\&12\catcode `\#12\catcode `\^12\catcode `\_12\catcode `\%12\relax}%
\providecommand \@@startlink[1]{}%
\providecommand \@@endlink[0]{}%
\providecommand \url  [0]{\begingroup\@sanitize@url \@url }%
\providecommand \@url [1]{\endgroup\@href {#1}{\urlprefix }}%
\providecommand \urlprefix  [0]{URL }%
\providecommand \Eprint [0]{\href }%
\providecommand \doibase [0]{http://dx.doi.org/}%
\providecommand \selectlanguage [0]{\@gobble}%
\providecommand \bibinfo  [0]{\@secondoftwo}%
\providecommand \bibfield  [0]{\@secondoftwo}%
\providecommand \translation [1]{[#1]}%
\providecommand \BibitemOpen [0]{}%
\providecommand \bibitemStop [0]{}%
\providecommand \bibitemNoStop [0]{.\EOS\space}%
\providecommand \EOS [0]{\spacefactor3000\relax}%
\providecommand \BibitemShut  [1]{\csname bibitem#1\endcsname}%
\let\auto@bib@innerbib\@empty
\bibitem [{\citenamefont {Herlach}\ \emph {et~al.}(2007)\citenamefont
  {Herlach}, \citenamefont {Galenko},\ and\ \citenamefont
  {Holland-Moritz}}]{Herlach.2007}%
  \BibitemOpen
  \bibfield  {author} {\bibinfo {author} {\bibfnamefont {D.~M.}\ \bibnamefont
  {Herlach}}, \bibinfo {author} {\bibfnamefont {P.~K.}\ \bibnamefont
  {Galenko}}, \ and\ \bibinfo {author} {\bibfnamefont {D.}~\bibnamefont
  {Holland-Moritz}},\ }\href@noop {} {\emph {\bibinfo {title} {Metastable
  solids from undercooled melts}}}\ (\bibinfo  {publisher} {Elsevier},\
  \bibinfo {address} {Amsterdam},\ \bibinfo {year} {2007})\BibitemShut
  {NoStop}%
\bibitem [{\citenamefont {{\relax Kelton}}\ and\ \citenamefont {{\relax
  Greer}}(2010)}]{Kelton.2010}%
  \BibitemOpen
  \bibfield  {author} {\bibinfo {author} {\bibfnamefont {K.~F.}\ \bibnamefont
  {{\relax Kelton}}}\ and\ \bibinfo {author} {\bibfnamefont {A.~L.}\
  \bibnamefont {{\relax Greer}}},\ }\href@noop {} {\emph {\bibinfo {title}
  {Nucleation in Condensed Matter: Applications in Materials and Biology}}}\
  (\bibinfo  {publisher} {Elsevier},\ \bibinfo {address} {Amsterdam},\ \bibinfo
  {year} {2010})\BibitemShut {NoStop}%
\bibitem [{\citenamefont {{\relax Yan}}\ \emph {et~al.}(2016)\citenamefont
  {{\relax Yan}}, \citenamefont {{\relax Song}}, \citenamefont {{\relax Hu}},
  \citenamefont {{\relax Dai}}, \citenamefont {{\relax Chu}},\ and\
  \citenamefont {{\relax Eckert}}}]{Yan.2016}%
  \BibitemOpen
  \bibfield  {author} {\bibinfo {author} {\bibfnamefont {Z.}~\bibnamefont
  {{\relax Yan}}}, \bibinfo {author} {\bibfnamefont {K.}~\bibnamefont {{\relax
  Song}}}, \bibinfo {author} {\bibfnamefont {Y.}~\bibnamefont {{\relax Hu}}},
  \bibinfo {author} {\bibfnamefont {F.}~\bibnamefont {{\relax Dai}}}, \bibinfo
  {author} {\bibfnamefont {Z.}~\bibnamefont {{\relax Chu}}}, \ and\ \bibinfo
  {author} {\bibfnamefont {J.}~\bibnamefont {{\relax Eckert}}},\ }\href@noop {}
  {\bibfield  {journal} {\bibinfo  {journal} {Sci. Rep.}\ }\textbf {\bibinfo
  {volume} {6}},\ \bibinfo {pages} {19358} (\bibinfo {year}
  {2016})}\BibitemShut {NoStop}%
\bibitem [{\citenamefont {{\relax Diao}}\ \emph {et~al.}(2015)\citenamefont
  {{\relax Diao}}, \citenamefont {{\relax Zhou}}, \citenamefont {{\relax
  Kurosawa}}, \citenamefont {{\relax Shaw}}, \citenamefont {{\relax Wang}},
  \citenamefont {{\relax Park}}, \citenamefont {{\relax Guo}}, \citenamefont
  {{\relax Reinspach}}, \citenamefont {{\relax Gu}}, \citenamefont {{\relax
  Gu}}, \citenamefont {{\relax Tee}}, \citenamefont {{\relax Pang}},
  \citenamefont {{\relax Yan}}, \citenamefont {{\relax Zhao}}, \citenamefont
  {{\relax Toney}}, \citenamefont {{\relax Mannsfeld}},\ and\ \citenamefont
  {{\relax Bao}}}]{Diao.2015}%
  \BibitemOpen
  \bibfield  {author} {\bibinfo {author} {\bibfnamefont {Y.}~\bibnamefont
  {{\relax Diao}}}, \bibinfo {author} {\bibfnamefont {Y.}~\bibnamefont {{\relax
  Zhou}}}, \bibinfo {author} {\bibfnamefont {T.}~\bibnamefont {{\relax
  Kurosawa}}}, \bibinfo {author} {\bibfnamefont {L.}~\bibnamefont {{\relax
  Shaw}}}, \bibinfo {author} {\bibfnamefont {C.}~\bibnamefont {{\relax Wang}}},
  \bibinfo {author} {\bibfnamefont {S.}~\bibnamefont {{\relax Park}}}, \bibinfo
  {author} {\bibfnamefont {Y.}~\bibnamefont {{\relax Guo}}}, \bibinfo {author}
  {\bibfnamefont {J.~A.}\ \bibnamefont {{\relax Reinspach}}}, \bibinfo {author}
  {\bibfnamefont {K.}~\bibnamefont {{\relax Gu}}}, \bibinfo {author}
  {\bibfnamefont {X.}~\bibnamefont {{\relax Gu}}}, \bibinfo {author}
  {\bibfnamefont {B.~C.~K.}\ \bibnamefont {{\relax Tee}}}, \bibinfo {author}
  {\bibfnamefont {C.}~\bibnamefont {{\relax Pang}}}, \bibinfo {author}
  {\bibfnamefont {H.}~\bibnamefont {{\relax Yan}}}, \bibinfo {author}
  {\bibfnamefont {D.}~\bibnamefont {{\relax Zhao}}}, \bibinfo {author}
  {\bibfnamefont {M.~F.}\ \bibnamefont {{\relax Toney}}}, \bibinfo {author}
  {\bibfnamefont {S.~C.~B.}\ \bibnamefont {{\relax Mannsfeld}}}, \ and\
  \bibinfo {author} {\bibfnamefont {Z.}~\bibnamefont {{\relax Bao}}},\
  }\href@noop {} {\bibfield  {journal} {\bibinfo  {journal} {Nature Commun.}\
  }\textbf {\bibinfo {volume} {6}},\ \bibinfo {pages} {7955} (\bibinfo {year}
  {2015})}\BibitemShut {NoStop}%
\bibitem [{\citenamefont {{\relax Anwar}}\ \emph {et~al.}(2014)\citenamefont
  {{\relax Anwar}}, \citenamefont {{\relax Berryman}},\ and\ \citenamefont
  {{\relax Schilling}}}]{Anwar.2014}%
  \BibitemOpen
  \bibfield  {author} {\bibinfo {author} {\bibfnamefont {M.}~\bibnamefont
  {{\relax Anwar}}}, \bibinfo {author} {\bibfnamefont {J.~T.}\ \bibnamefont
  {{\relax Berryman}}}, \ and\ \bibinfo {author} {\bibfnamefont
  {T.}~\bibnamefont {{\relax Schilling}}},\ }\href@noop {} {\bibfield
  {journal} {\bibinfo  {journal} {J. Chem. Phys.}\ }\textbf {\bibinfo {volume}
  {141}},\ \bibinfo {pages} {124910} (\bibinfo {year} {2014})}\BibitemShut
  {NoStop}%
\bibitem [{\citenamefont {{\relax Bokeloh}}\ \emph {et~al.}(2011)\citenamefont
  {{\relax Bokeloh}}, \citenamefont {{\relax Rozas}}, \citenamefont {{\relax
  Horbach}},\ and\ \citenamefont {{\relax Wilde}}}]{Bokeloh.2011}%
  \BibitemOpen
  \bibfield  {author} {\bibinfo {author} {\bibfnamefont {J.}~\bibnamefont
  {{\relax Bokeloh}}}, \bibinfo {author} {\bibfnamefont {R.~E.}\ \bibnamefont
  {{\relax Rozas}}}, \bibinfo {author} {\bibfnamefont {J.}~\bibnamefont
  {{\relax Horbach}}}, \ and\ \bibinfo {author} {\bibfnamefont
  {G.}~\bibnamefont {{\relax Wilde}}},\ }\href@noop {} {\bibfield  {journal}
  {\bibinfo  {journal} {Phys. Rev. Lett.}\ }\textbf {\bibinfo {volume} {107}},\
  \bibinfo {pages} {145701} (\bibinfo {year} {2011})}\BibitemShut {NoStop}%
\bibitem [{\citenamefont {{\relax Richard}}\ and\ \citenamefont {{\relax
  Speck}}(2015)}]{Richard.2015}%
  \BibitemOpen
  \bibfield  {author} {\bibinfo {author} {\bibfnamefont {D.}~\bibnamefont
  {{\relax Richard}}}\ and\ \bibinfo {author} {\bibfnamefont {T.}~\bibnamefont
  {{\relax Speck}}},\ }\href@noop {} {\bibfield  {journal} {\bibinfo  {journal}
  {Sci. Rep.}\ }\textbf {\bibinfo {volume} {5}},\ \bibinfo {pages} {14610}
  (\bibinfo {year} {2015})}\BibitemShut {NoStop}%
\bibitem [{\citenamefont {{\relax Wu}}(2009)}]{Wu.2009}%
  \BibitemOpen
  \bibfield  {author} {\bibinfo {author} {\bibfnamefont {Y.~L.}\ \bibnamefont
  {{\relax Wu}}},\ }\href@noop {} {\bibfield  {journal} {\bibinfo  {journal}
  {Proc. Natl. Acad. Sci. USA}\ }\textbf {\bibinfo {volume} {106}},\ \bibinfo
  {pages} {10564} (\bibinfo {year} {2009})}\BibitemShut {NoStop}%
\bibitem [{\citenamefont {{\relax Cerd\`a}}\ \emph {et~al.}(2008)\citenamefont
  {{\relax Cerd\`a}}, \citenamefont {{\relax Sintes}}, \citenamefont {{\relax
  Holm}}, \citenamefont {{\relax Sorensen}},\ and\ \citenamefont {{\relax
  Chakrabarti}}}]{Cerda.2008}%
  \BibitemOpen
  \bibfield  {author} {\bibinfo {author} {\bibfnamefont {J.~J.}\ \bibnamefont
  {{\relax Cerd\`a}}}, \bibinfo {author} {\bibfnamefont {T.}~\bibnamefont
  {{\relax Sintes}}}, \bibinfo {author} {\bibfnamefont {C.}~\bibnamefont
  {{\relax Holm}}}, \bibinfo {author} {\bibfnamefont {C.~M.}\ \bibnamefont
  {{\relax Sorensen}}}, \ and\ \bibinfo {author} {\bibfnamefont
  {A.}~\bibnamefont {{\relax Chakrabarti}}},\ }\href@noop {} {\bibfield
  {journal} {\bibinfo  {journal} {Phys. Rev. E}\ }\textbf {\bibinfo {volume}
  {78}},\ \bibinfo {pages} {031403} (\bibinfo {year} {2008})}\BibitemShut
  {NoStop}%
\bibitem [{\citenamefont {{\relax Mokshin}}\ and\ \citenamefont {{\relax
  Barrat}}(2008)}]{Mokshin.2008}%
  \BibitemOpen
  \bibfield  {author} {\bibinfo {author} {\bibfnamefont {A.~V.}\ \bibnamefont
  {{\relax Mokshin}}}\ and\ \bibinfo {author} {\bibfnamefont {J.-L.}\
  \bibnamefont {{\relax Barrat}}},\ }\href@noop {} {\bibfield  {journal}
  {\bibinfo  {journal} {Phys. Rev. E}\ }\textbf {\bibinfo {volume} {77}},\
  \bibinfo {pages} {021505} (\bibinfo {year} {2008})}\BibitemShut {NoStop}%
\bibitem [{\citenamefont {{\relax Shao}}\ \emph {et~al.}(2015)\citenamefont
  {{\relax Shao}}, \citenamefont {{\relax Singer}}, \citenamefont {{\relax
  Liu}}, \citenamefont {{\relax Liu}}, \citenamefont {{\relax Li}},
  \citenamefont {{\relax Gopinadhan}}, \citenamefont {{\relax O'Hern}},
  \citenamefont {{\relax Schroers}},\ and\ \citenamefont {{\relax
  Osuji}}}]{Shao.2015}%
  \BibitemOpen
  \bibfield  {author} {\bibinfo {author} {\bibfnamefont {Z.}~\bibnamefont
  {{\relax Shao}}}, \bibinfo {author} {\bibfnamefont {J.~P.}\ \bibnamefont
  {{\relax Singer}}}, \bibinfo {author} {\bibfnamefont {Y.}~\bibnamefont
  {{\relax Liu}}}, \bibinfo {author} {\bibfnamefont {Z.}~\bibnamefont {{\relax
  Liu}}}, \bibinfo {author} {\bibfnamefont {H.}~\bibnamefont {{\relax Li}}},
  \bibinfo {author} {\bibfnamefont {M.}~\bibnamefont {{\relax Gopinadhan}}},
  \bibinfo {author} {\bibfnamefont {C.~S.}\ \bibnamefont {{\relax O'Hern}}},
  \bibinfo {author} {\bibfnamefont {J.}~\bibnamefont {{\relax Schroers}}}, \
  and\ \bibinfo {author} {\bibfnamefont {C.~O.}\ \bibnamefont {{\relax
  Osuji}}},\ }\href@noop {} {\bibfield  {journal} {\bibinfo  {journal} {Phys.
  Rev. E}\ }\textbf {\bibinfo {volume} {91}},\ \bibinfo {pages} {020301(R)}
  (\bibinfo {year} {2015})}\BibitemShut {NoStop}%
\bibitem [{\citenamefont {{\relax Blaak}}\ \emph {et~al.}(2004)\citenamefont
  {{\relax Blaak}}, \citenamefont {{\relax Auer}}, \citenamefont {{\relax
  Frenkel}},\ and\ \citenamefont {{\relax L\"owen}}}]{Blaak.2004}%
  \BibitemOpen
  \bibfield  {author} {\bibinfo {author} {\bibfnamefont {R.}~\bibnamefont
  {{\relax Blaak}}}, \bibinfo {author} {\bibfnamefont {S.}~\bibnamefont
  {{\relax Auer}}}, \bibinfo {author} {\bibfnamefont {D.}~\bibnamefont {{\relax
  Frenkel}}}, \ and\ \bibinfo {author} {\bibfnamefont {H.}~\bibnamefont
  {{\relax L\"owen}}},\ }\href@noop {} {\bibfield  {journal} {\bibinfo
  {journal} {Phys. Rev. Lett.}\ }\textbf {\bibinfo {volume} {93}},\ \bibinfo
  {pages} {068303} (\bibinfo {year} {2004})}\BibitemShut {NoStop}%
\bibitem [{\citenamefont {{\relax Mokshin}}\ \emph {et~al.}(2013)\citenamefont
  {{\relax Mokshin}}, \citenamefont {{\relax Galimzyanov}},\ and\ \citenamefont
  {{\relax Barrat}}}]{Mokshin.2013}%
  \BibitemOpen
  \bibfield  {author} {\bibinfo {author} {\bibfnamefont {A.~V.}\ \bibnamefont
  {{\relax Mokshin}}}, \bibinfo {author} {\bibfnamefont {B.~N.}\ \bibnamefont
  {{\relax Galimzyanov}}}, \ and\ \bibinfo {author} {\bibfnamefont {J.-L.}\
  \bibnamefont {{\relax Barrat}}},\ }\href@noop {} {\bibfield  {journal}
  {\bibinfo  {journal} {Phys. Rev. E}\ }\textbf {\bibinfo {volume} {87}},\
  \bibinfo {pages} {062307} (\bibinfo {year} {2013})}\BibitemShut {NoStop}%
\bibitem [{\citenamefont {{\relax Mura}}\ and\ \citenamefont {{\relax
  Zaccone}}(2016)}]{Mura.2016}%
  \BibitemOpen
  \bibfield  {author} {\bibinfo {author} {\bibfnamefont {F.}~\bibnamefont
  {{\relax Mura}}}\ and\ \bibinfo {author} {\bibfnamefont {A.}~\bibnamefont
  {{\relax Zaccone}}},\ }\href@noop {} {\bibfield  {journal} {\bibinfo
  {journal} {Phys. Rev. E}\ }\textbf {\bibinfo {volume} {93}},\ \bibinfo
  {pages} {042803} (\bibinfo {year} {2016})}\BibitemShut {NoStop}%
\bibitem [{\citenamefont {{\relax Lander}}\ \emph {et~al.}(2013)\citenamefont
  {{\relax Lander}}, \citenamefont {{\relax Seifert}},\ and\ \citenamefont
  {{\relax Speck}}}]{Lander.2013}%
  \BibitemOpen
  \bibfield  {author} {\bibinfo {author} {\bibfnamefont {B.}~\bibnamefont
  {{\relax Lander}}}, \bibinfo {author} {\bibfnamefont {U.}~\bibnamefont
  {{\relax Seifert}}}, \ and\ \bibinfo {author} {\bibfnamefont
  {T.}~\bibnamefont {{\relax Speck}}},\ }\href@noop {} {\bibfield  {journal}
  {\bibinfo  {journal} {J. Chem. Phys.}\ }\textbf {\bibinfo {volume} {138}},\
  \bibinfo {pages} {224907} (\bibinfo {year} {2013})}\BibitemShut {NoStop}%
\bibitem [{\citenamefont {{\relax Dorosz}}\ \emph {et~al.}(2016)\citenamefont
  {{\relax Dorosz}}, \citenamefont {{\relax Voigtmann}},\ and\ \citenamefont
  {{\relax Schilling}}}]{Dorosz.2016}%
  \BibitemOpen
  \bibfield  {author} {\bibinfo {author} {\bibfnamefont {S.}~\bibnamefont
  {{\relax Dorosz}}}, \bibinfo {author} {\bibfnamefont {{\relax
  Th.}.}~\bibnamefont {{\relax Voigtmann}}}, \ and\ \bibinfo {author}
  {\bibfnamefont {T.}~\bibnamefont {{\relax Schilling}}},\ }\href@noop {}
  {\bibfield  {journal} {\bibinfo  {journal} {EPL}\ }\textbf {\bibinfo {volume}
  {113}},\ \bibinfo {pages} {10004} (\bibinfo {year} {2016})}\BibitemShut
  {NoStop}%
\bibitem [{\citenamefont {{\relax Butler}}\ and\ \citenamefont {{\relax
  Harrowell}}(2002)}]{Butler.2002}%
  \BibitemOpen
  \bibfield  {author} {\bibinfo {author} {\bibfnamefont {S.}~\bibnamefont
  {{\relax Butler}}}\ and\ \bibinfo {author} {\bibfnamefont {P.}~\bibnamefont
  {{\relax Harrowell}}},\ }\href@noop {} {\bibfield  {journal} {\bibinfo
  {journal} {Nature (London)}\ }\textbf {\bibinfo {volume} {415}},\ \bibinfo
  {pages} {1008} (\bibinfo {year} {2002})}\BibitemShut {NoStop}%
\bibitem [{\citenamefont {{\relax Butler}}\ and\ \citenamefont {{\relax
  Harrowell}}(2003{\natexlab{a}})}]{Butler.2003}%
  \BibitemOpen
  \bibfield  {author} {\bibinfo {author} {\bibfnamefont {S.}~\bibnamefont
  {{\relax Butler}}}\ and\ \bibinfo {author} {\bibfnamefont {P.}~\bibnamefont
  {{\relax Harrowell}}},\ }\href@noop {} {\bibfield  {journal} {\bibinfo
  {journal} {J. Chem. Phys.}\ }\textbf {\bibinfo {volume} {118}},\ \bibinfo
  {pages} {4115} (\bibinfo {year} {2003}{\natexlab{a}})}\BibitemShut {NoStop}%
\bibitem [{\citenamefont {{\relax Butler}}\ and\ \citenamefont {{\relax
  Harrowell}}(2003{\natexlab{b}})}]{Butler.2003b}%
  \BibitemOpen
  \bibfield  {author} {\bibinfo {author} {\bibfnamefont {S.}~\bibnamefont
  {{\relax Butler}}}\ and\ \bibinfo {author} {\bibfnamefont {P.}~\bibnamefont
  {{\relax Harrowell}}},\ }\href@noop {} {\bibfield  {journal} {\bibinfo
  {journal} {Phys. Rev. E}\ }\textbf {\bibinfo {volume} {67}},\ \bibinfo
  {pages} {051503} (\bibinfo {year} {2003}{\natexlab{b}})}\BibitemShut
  {NoStop}%
\bibitem [{\citenamefont {{\relax Mishin}}\ \emph {et~al.}(2002)\citenamefont
  {{\relax Mishin}}, \citenamefont {{\relax Mehl}},\ and\ \citenamefont
  {{\relax Papaconstantopoulos}}}]{Mishin.2002}%
  \BibitemOpen
  \bibfield  {author} {\bibinfo {author} {\bibfnamefont {Y.}~\bibnamefont
  {{\relax Mishin}}}, \bibinfo {author} {\bibfnamefont {M.~J.}\ \bibnamefont
  {{\relax Mehl}}}, \ and\ \bibinfo {author} {\bibfnamefont {D.~A.}\
  \bibnamefont {{\relax Papaconstantopoulos}}},\ }\href@noop {} {\bibfield
  {journal} {\bibinfo  {journal} {Phys. Rev. B}\ }\textbf {\bibinfo {volume}
  {65}},\ \bibinfo {pages} {224114} (\bibinfo {year} {2002})}\BibitemShut
  {NoStop}%
\bibitem [{\citenamefont {Das}\ \emph {et~al.}(2005)\citenamefont {Das},
  \citenamefont {Horbach}, \citenamefont {Koza}, \citenamefont {Chatoth},\ and\
  \citenamefont {Meyer}}]{Das.2005}%
  \BibitemOpen
  \bibfield  {author} {\bibinfo {author} {\bibfnamefont {S.~K.}\ \bibnamefont
  {Das}}, \bibinfo {author} {\bibfnamefont {J.}~\bibnamefont {Horbach}},
  \bibinfo {author} {\bibfnamefont {M.~M.}\ \bibnamefont {Koza}}, \bibinfo
  {author} {\bibfnamefont {S.~M.}\ \bibnamefont {Chatoth}}, \ and\ \bibinfo
  {author} {\bibfnamefont {A.}~\bibnamefont {Meyer}},\ }\href@noop {}
  {\bibfield  {journal} {\bibinfo  {journal} {Appl. Phys. Lett.}\ }\textbf
  {\bibinfo {volume} {86}},\ \bibinfo {pages} {011918} (\bibinfo {year}
  {2005})}\BibitemShut {NoStop}%
\bibitem [{\citenamefont {{\relax Tang}}\ and\ \citenamefont {{\relax
  Harrowell}}(2013)}]{Tang.2013}%
  \BibitemOpen
  \bibfield  {author} {\bibinfo {author} {\bibfnamefont {C.}~\bibnamefont
  {{\relax Tang}}}\ and\ \bibinfo {author} {\bibfnamefont {P.}~\bibnamefont
  {{\relax Harrowell}}},\ }\href@noop {} {\bibfield  {journal} {\bibinfo
  {journal} {Nature Materials}\ }\textbf {\bibinfo {volume} {12}},\ \bibinfo
  {pages} {507} (\bibinfo {year} {2013})}\BibitemShut {NoStop}%
\bibitem [{\citenamefont {{\relax Kerrache}}\ \emph {et~al.}(2008)\citenamefont
  {{\relax Kerrache}}, \citenamefont {{\relax Horbach}},\ and\ \citenamefont
  {{\relax Binder}}}]{Kerrache.2008}%
  \BibitemOpen
  \bibfield  {author} {\bibinfo {author} {\bibfnamefont {A.}~\bibnamefont
  {{\relax Kerrache}}}, \bibinfo {author} {\bibfnamefont {J.}~\bibnamefont
  {{\relax Horbach}}}, \ and\ \bibinfo {author} {\bibfnamefont
  {K.}~\bibnamefont {{\relax Binder}}},\ }\href@noop {} {\bibfield  {journal}
  {\bibinfo  {journal} {EPL}\ }\textbf {\bibinfo {volume} {81}},\ \bibinfo
  {pages} {58001} (\bibinfo {year} {2008})}\BibitemShut {NoStop}%
\bibitem [{\citenamefont {{\relax Kuhn}}\ and\ \citenamefont {{\relax
  Horbach}}(2013)}]{Kuhn.2013}%
  \BibitemOpen
  \bibfield  {author} {\bibinfo {author} {\bibfnamefont {P.}~\bibnamefont
  {{\relax Kuhn}}}\ and\ \bibinfo {author} {\bibfnamefont {J.}~\bibnamefont
  {{\relax Horbach}}},\ }\href@noop {} {\bibfield  {journal} {\bibinfo
  {journal} {Phys. Rev. B}\ }\textbf {\bibinfo {volume} {87}},\ \bibinfo
  {pages} {014105} (\bibinfo {year} {2013})}\BibitemShut {NoStop}%
\bibitem [{\citenamefont {{\relax Zheng}}\ \emph {et~al.}(2012)\citenamefont
  {{\relax Zheng}}, \citenamefont {{\relax Yang}}, \citenamefont {{\relax
  Gao}}, \citenamefont {{\relax Hoyt}}, \citenamefont {{\relax Asta}},\ and\
  \citenamefont {{\relax Sun}}}]{Zheng.2012}%
  \BibitemOpen
  \bibfield  {author} {\bibinfo {author} {\bibfnamefont {X.~Q.}\ \bibnamefont
  {{\relax Zheng}}}, \bibinfo {author} {\bibfnamefont {Y.}~\bibnamefont
  {{\relax Yang}}}, \bibinfo {author} {\bibfnamefont {Y.~F.}\ \bibnamefont
  {{\relax Gao}}}, \bibinfo {author} {\bibfnamefont {J.~J.}\ \bibnamefont
  {{\relax Hoyt}}}, \bibinfo {author} {\bibfnamefont {M.}~\bibnamefont {{\relax
  Asta}}}, \ and\ \bibinfo {author} {\bibfnamefont {D.~Y.}\ \bibnamefont
  {{\relax Sun}}},\ }\href@noop {} {\bibfield  {journal} {\bibinfo  {journal}
  {Phys. Rev. E}\ }\textbf {\bibinfo {volume} {85}},\ \bibinfo {pages} {041601}
  (\bibinfo {year} {2012})}\BibitemShut {NoStop}%
\bibitem [{\citenamefont {{\relax Evans}}\ and\ \citenamefont {{\relax
  Morriss}}(2008)}]{Evans}%
  \BibitemOpen
  \bibfield  {author} {\bibinfo {author} {\bibfnamefont {D.~J.}\ \bibnamefont
  {{\relax Evans}}}\ and\ \bibinfo {author} {\bibfnamefont {G.~P.}\
  \bibnamefont {{\relax Morriss}}},\ }\href@noop {} {\emph {\bibinfo {title}
  {Statistical Mechanics of Nonequilibrium Liquids}}},\ \bibinfo {edition}
  {2nd}\ ed.\ (\bibinfo  {publisher} {Cambridge University Press},\ \bibinfo
  {year} {2008})\BibitemShut {NoStop}%
\bibitem [{\citenamefont {{\relax Monk}}\ \emph {et~al.}(2010)\citenamefont
  {{\relax Monk}}, \citenamefont {{\relax Yang}}, \citenamefont {{\relax
  Mendelev}}, \citenamefont {{\relax Asta}}, \citenamefont {{\relax Hoyt}},\
  and\ \citenamefont {{\relax Sun}}}]{Monk.2010}%
  \BibitemOpen
  \bibfield  {author} {\bibinfo {author} {\bibfnamefont {J.}~\bibnamefont
  {{\relax Monk}}}, \bibinfo {author} {\bibfnamefont {Y.}~\bibnamefont {{\relax
  Yang}}}, \bibinfo {author} {\bibfnamefont {M.~I.}\ \bibnamefont {{\relax
  Mendelev}}}, \bibinfo {author} {\bibfnamefont {M.}~\bibnamefont {{\relax
  Asta}}}, \bibinfo {author} {\bibfnamefont {J.~J.}\ \bibnamefont {{\relax
  Hoyt}}}, \ and\ \bibinfo {author} {\bibfnamefont {D.~Y.}\ \bibnamefont
  {{\relax Sun}}},\ }\href@noop {} {\bibfield  {journal} {\bibinfo  {journal}
  {Modell. Simul. Mater. Sci. Eng.}\ }\textbf {\bibinfo {volume} {18}},\
  \bibinfo {pages} {015004} (\bibinfo {year} {2010})}\BibitemShut {NoStop}%
\bibitem [{Note1()}]{Note1}%
  \BibitemOpen
  \bibinfo {note} {The time scales of single-particle vibrations, $\tau _0$,
  and that of collective structural relaxation, $\tau $, have been determined
  from the intermediate scattering functions in the quiescent fluid, at wave
  vectors corresponding to typical interparticle distances.}\BibitemShut
  {Stop}%
\bibitem [{\citenamefont {{\relax Herlach}}(2014)}]{Herlach.2014}%
  \BibitemOpen
  \bibfield  {author} {\bibinfo {author} {\bibfnamefont {D.~M.}\ \bibnamefont
  {{\relax Herlach}}},\ }\href@noop {} {\bibfield  {journal} {\bibinfo
  {journal} {Metals}\ }\textbf {\bibinfo {volume} {4}},\ \bibinfo {pages} {196}
  (\bibinfo {year} {2014})}\BibitemShut {NoStop}%
\bibitem [{\citenamefont {{\relax Wang}}\ \emph {et~al.}(2014)\citenamefont
  {{\relax Wang}}, \citenamefont {{\relax Herlach}},\ and\ \citenamefont
  {{\relax Liu}}}]{Wang.2014}%
  \BibitemOpen
  \bibfield  {author} {\bibinfo {author} {\bibfnamefont {H.}~\bibnamefont
  {{\relax Wang}}}, \bibinfo {author} {\bibfnamefont {D.~M.}\ \bibnamefont
  {{\relax Herlach}}}, \ and\ \bibinfo {author} {\bibfnamefont {R.~P.}\
  \bibnamefont {{\relax Liu}}},\ }\href@noop {} {\bibfield  {journal} {\bibinfo
   {journal} {EPL}\ }\textbf {\bibinfo {volume} {105}},\ \bibinfo {pages}
  {36001} (\bibinfo {year} {2014})}\BibitemShut {NoStop}%
\bibitem [{\citenamefont {{\relax Guan}}\ \emph {et~al.}(2010)\citenamefont
  {{\relax Guan}}, \citenamefont {{\relax Chen}},\ and\ \citenamefont {{\relax
  Egami}}}]{Guan.2010}%
  \BibitemOpen
  \bibfield  {author} {\bibinfo {author} {\bibfnamefont {P.}~\bibnamefont
  {{\relax Guan}}}, \bibinfo {author} {\bibfnamefont {M.}~\bibnamefont {{\relax
  Chen}}}, \ and\ \bibinfo {author} {\bibfnamefont {T.}~\bibnamefont {{\relax
  Egami}}},\ }\href@noop {} {\bibfield  {journal} {\bibinfo  {journal} {Phys.
  Rev. Lett.}\ }\textbf {\bibinfo {volume} {104}},\ \bibinfo {pages} {205701}
  (\bibinfo {year} {2010})}\BibitemShut {NoStop}%
\bibitem [{\citenamefont {{\relax Voigtmann}}(2014)}]{Voigtmann.2014}%
  \BibitemOpen
  \bibfield  {author} {\bibinfo {author} {\bibfnamefont {{\relax
  Th}.}~\bibnamefont {{\relax Voigtmann}}},\ }\href@noop {} {\bibfield
  {journal} {\bibinfo  {journal} {Curr. Opin. Colloid Interf. Sci.}\ }\textbf
  {\bibinfo {volume} {19}},\ \bibinfo {pages} {549} (\bibinfo {year}
  {2014})}\BibitemShut {NoStop}%
\bibitem [{\citenamefont {{\relax Ramsay}}\ and\ \citenamefont {{\relax
  Harrowell}}(2016)}]{Ramsay.2016}%
  \BibitemOpen
  \bibfield  {author} {\bibinfo {author} {\bibfnamefont {M.}~\bibnamefont
  {{\relax Ramsay}}}\ and\ \bibinfo {author} {\bibfnamefont {P.}~\bibnamefont
  {{\relax Harrowell}}},\ }\href@noop {} {\bibfield  {journal} {\bibinfo
  {journal} {Phys. Rev. E}\ }\textbf {\bibinfo {volume} {93}},\ \bibinfo
  {pages} {042608} (\bibinfo {year} {2016})}\BibitemShut {NoStop}%
\bibitem [{\citenamefont {{\relax Berthier}}\ and\ \citenamefont {{\relax
  Barrat}}(2002)}]{Berthier.2002}%
  \BibitemOpen
  \bibfield  {author} {\bibinfo {author} {\bibfnamefont {L.}~\bibnamefont
  {{\relax Berthier}}}\ and\ \bibinfo {author} {\bibfnamefont {J.-L.}\
  \bibnamefont {{\relax Barrat}}},\ }\href@noop {} {\bibfield  {journal}
  {\bibinfo  {journal} {Phys. Rev. Lett.}\ }\textbf {\bibinfo {volume} {89}},\
  \bibinfo {pages} {095702} (\bibinfo {year} {2002})}\BibitemShut {NoStop}%
\bibitem [{\citenamefont {Lee}\ \emph {et~al.}(2014)\citenamefont {Lee},
  \citenamefont {Matson}, \citenamefont {Binder}, \citenamefont {Kolbe},
  \citenamefont {Herlach},\ and\ \citenamefont {Hyers}}]{Lee.2014}%
  \BibitemOpen
  \bibfield  {author} {\bibinfo {author} {\bibfnamefont {J.}~\bibnamefont
  {Lee}}, \bibinfo {author} {\bibfnamefont {D.~M.}\ \bibnamefont {Matson}},
  \bibinfo {author} {\bibfnamefont {S.}~\bibnamefont {Binder}}, \bibinfo
  {author} {\bibfnamefont {M.}~\bibnamefont {Kolbe}}, \bibinfo {author}
  {\bibfnamefont {D.}~\bibnamefont {Herlach}}, \ and\ \bibinfo {author}
  {\bibfnamefont {R.~W.}\ \bibnamefont {Hyers}},\ }\href@noop {} {\bibfield
  {journal} {\bibinfo  {journal} {Metall. Mater. Trans. B}\ }\textbf {\bibinfo
  {volume} {45}},\ \bibinfo {pages} {1018} (\bibinfo {year}
  {2014})}\BibitemShut {NoStop}%
\bibitem [{\citenamefont {{\relax Shevchenko}}\ \emph
  {et~al.}(2015)\citenamefont {{\relax Shevchenko}}, \citenamefont {{\relax
  Roshchupkina}}, \citenamefont {{\relax Sokolova}},\ and\ \citenamefont
  {{\relax Eckert}}}]{Shevchenko.2015}%
  \BibitemOpen
  \bibfield  {author} {\bibinfo {author} {\bibfnamefont {N.}~\bibnamefont
  {{\relax Shevchenko}}}, \bibinfo {author} {\bibfnamefont {O.}~\bibnamefont
  {{\relax Roshchupkina}}}, \bibinfo {author} {\bibfnamefont {O.}~\bibnamefont
  {{\relax Sokolova}}}, \ and\ \bibinfo {author} {\bibfnamefont
  {S.}~\bibnamefont {{\relax Eckert}}},\ }\href@noop {} {\bibfield  {journal}
  {\bibinfo  {journal} {J. Cryst. Growth}\ }\textbf {\bibinfo {volume} {417}},\
  \bibinfo {pages} {1} (\bibinfo {year} {2015})}\BibitemShut {NoStop}%
\bibitem [{\citenamefont {{\relax Rojas}}\ \emph {et~al.}(2015)\citenamefont
  {{\relax Rojas}}, \citenamefont {{\relax Takaki}},\ and\ \citenamefont
  {{\relax Ohno}}}]{Rojas.2015}%
  \BibitemOpen
  \bibfield  {author} {\bibinfo {author} {\bibfnamefont {R.}~\bibnamefont
  {{\relax Rojas}}}, \bibinfo {author} {\bibfnamefont {T.}~\bibnamefont
  {{\relax Takaki}}}, \ and\ \bibinfo {author} {\bibfnamefont {M.}~\bibnamefont
  {{\relax Ohno}}},\ }\href@noop {} {\bibfield  {journal} {\bibinfo  {journal}
  {J. Comput. Phys.}\ }\textbf {\bibinfo {volume} {298}},\ \bibinfo {pages}
  {29} (\bibinfo {year} {2015})}\BibitemShut {NoStop}%
\bibitem [{\citenamefont {{\relax Reutzel}}\ \emph {et~al.}(2007)\citenamefont
  {{\relax Reutzel}}, \citenamefont {{\relax Hartmann}}, \citenamefont {{\relax
  Galenko}}, \citenamefont {{\relax Schneider}},\ and\ \citenamefont {{\relax
  Herlach}}}]{Reutzel.2007}%
  \BibitemOpen
  \bibfield  {author} {\bibinfo {author} {\bibfnamefont {S.}~\bibnamefont
  {{\relax Reutzel}}}, \bibinfo {author} {\bibfnamefont {H.}~\bibnamefont
  {{\relax Hartmann}}}, \bibinfo {author} {\bibfnamefont {P.~K.}\ \bibnamefont
  {{\relax Galenko}}}, \bibinfo {author} {\bibfnamefont {S.}~\bibnamefont
  {{\relax Schneider}}}, \ and\ \bibinfo {author} {\bibfnamefont {D.~M.}\
  \bibnamefont {{\relax Herlach}}},\ }\href@noop {} {\bibfield  {journal}
  {\bibinfo  {journal} {Appl. Phys. Lett.}\ }\textbf {\bibinfo {volume} {91}},\
  \bibinfo {pages} {041913} (\bibinfo {year} {2007})}\BibitemShut {NoStop}%
\bibitem [{\citenamefont {{\relax Binder}}\ \emph {et~al.}(2014)\citenamefont
  {{\relax Binder}}, \citenamefont {{\relax Galenko}},\ and\ \citenamefont
  {{\relax Herlach}}}]{Binder.2014}%
  \BibitemOpen
  \bibfield  {author} {\bibinfo {author} {\bibfnamefont {S.}~\bibnamefont
  {{\relax Binder}}}, \bibinfo {author} {\bibfnamefont {P.~K.}\ \bibnamefont
  {{\relax Galenko}}}, \ and\ \bibinfo {author} {\bibfnamefont {D.~M.}\
  \bibnamefont {{\relax Herlach}}},\ }\href@noop {} {\bibfield  {journal}
  {\bibinfo  {journal} {J. Appl. Phys.}\ }\textbf {\bibinfo {volume} {115}},\
  \bibinfo {pages} {053511} (\bibinfo {year} {2014})}\BibitemShut {NoStop}%
\bibitem [{\citenamefont {{\relax Yang}}\ \emph {et~al.}(2011)\citenamefont
  {{\relax Yang}}, \citenamefont {{\relax Humadi}}, \citenamefont {{\relax
  Buta}}, \citenamefont {{\relax Laird}}, \citenamefont {{\relax Sun}},
  \citenamefont {{\relax Hoyt}},\ and\ \citenamefont {{\relax
  Asta}}}]{Yang.2011}%
  \BibitemOpen
  \bibfield  {author} {\bibinfo {author} {\bibfnamefont {Y.}~\bibnamefont
  {{\relax Yang}}}, \bibinfo {author} {\bibfnamefont {H.}~\bibnamefont {{\relax
  Humadi}}}, \bibinfo {author} {\bibfnamefont {D.}~\bibnamefont {{\relax
  Buta}}}, \bibinfo {author} {\bibfnamefont {B.~B.}\ \bibnamefont {{\relax
  Laird}}}, \bibinfo {author} {\bibfnamefont {D.}~\bibnamefont {{\relax Sun}}},
  \bibinfo {author} {\bibfnamefont {J.~J.}\ \bibnamefont {{\relax Hoyt}}}, \
  and\ \bibinfo {author} {\bibfnamefont {M.}~\bibnamefont {{\relax Asta}}},\
  }\href@noop {} {\bibfield  {journal} {\bibinfo  {journal} {Phys. Rev. Lett.}\
  }\textbf {\bibinfo {volume} {107}},\ \bibinfo {pages} {025505} (\bibinfo
  {year} {2011})}\BibitemShut {NoStop}%
\bibitem [{\citenamefont {Hartmann}\ \emph {et~al.}(2009)\citenamefont
  {Hartmann}, \citenamefont {Holland-Moritz}, \citenamefont {Galenko},\ and\
  \citenamefont {Herlach}}]{Hartmann.2009}%
  \BibitemOpen
  \bibfield  {author} {\bibinfo {author} {\bibfnamefont {H.}~\bibnamefont
  {Hartmann}}, \bibinfo {author} {\bibfnamefont {D.}~\bibnamefont
  {Holland-Moritz}}, \bibinfo {author} {\bibfnamefont {P.~K.}\ \bibnamefont
  {Galenko}}, \ and\ \bibinfo {author} {\bibfnamefont {D.~M.}\ \bibnamefont
  {Herlach}},\ }\href@noop {} {\bibfield  {journal} {\bibinfo  {journal} {EPL}\
  }\textbf {\bibinfo {volume} {87}},\ \bibinfo {pages} {40007} (\bibinfo {year}
  {2009})}\BibitemShut {NoStop}%
\bibitem [{\citenamefont {Lengsdorf}\ \emph {et~al.}(2010)\citenamefont
  {Lengsdorf}, \citenamefont {Holland-Moritz},\ and\ \citenamefont
  {Herlach}}]{Lengsdorf2010}%
  \BibitemOpen
  \bibfield  {author} {\bibinfo {author} {\bibfnamefont {R.}~\bibnamefont
  {Lengsdorf}}, \bibinfo {author} {\bibfnamefont {D.}~\bibnamefont
  {Holland-Moritz}}, \ and\ \bibinfo {author} {\bibfnamefont {D.~M.}\
  \bibnamefont {Herlach}},\ }\href@noop {} {\bibfield  {journal} {\bibinfo
  {journal} {Scr. Mater.}\ }\textbf {\bibinfo {volume} {62}},\ \bibinfo {pages}
  {365} (\bibinfo {year} {2010})}\BibitemShut {NoStop}%
\end{thebibliography}%
\bibliographystyle{apsrev4-1}

\end{document}